\documentstyle[12pt,epsf]{article}
\newcommand{\be}{\begin{equation}}
\newcommand{\ee}{\end{equation}}
\newcommand{\bd}{\begin{displaymath}}
\newcommand{\ed}{\end{displaymath}}
\newcommand{\baa}{\begin{array}{lll}}
\newcommand{\eaa}{\end{array}}
\newcommand{\ba}{\begin{eqnarray}}
\newcommand{\ea}{\end{eqnarray}}

\begin{document}
\begin{flushright}
CPT--2000/P.4003.
\end{flushright}

\begin{center}
{\Large\bf Positivity constraints  and flavor dependence
 of higher twists} \
\\[10mm]

{\bf \large J. Soffer \footnote{E-mail: soffer@cpt.univ-mrs.fr
}}\\

{\it Centre de Physique Th\'eorique - CNRS - Luminy,\\
Case 907 F-13288 Marseille Cedex 9 - France} \\

{ \bf \large and \ \ O. V. Teryaev\footnote{
E-mail: teryaev@thsun1.jinr.ru}} \\

{\it Bogoliubov Laboratory of Theoretical Physics, \\
Joint Institute for Nuclear Research, Dubna, 141980, Russia}
\end{center}
\begin{abstract}
The
positivity bound for the transverse asymmetry $A_2$ may be improved
and applied for each quark flavor separately.
We use it to test the consistency of
twist-two approximation for the transverse spin-dependent ($g_T$) and
the longitudinal spin-averaged ($F_L$) structure functions.
While it is satisfied in the case of
$u$ quarks, it might be violated for
$d$ quarks in a region of moderate $x$ because of its
negative polarization.
We attribute this inconsistency to a stronger twist-three
contribution, whose existence found by the QCD sum rule method
is a long-standing puzzle.


\end{abstract}


Positivity is playing a very important role in constraining various spin- \\
dependent observables, in particular by providing a bound for the
transverse asymmetry in polarized Deep Inelastic Scattering (DIS).
It is a well-known condition established long time ago \cite{CL}
and based on an extensive study by
Doncel and de Rafael \cite{DDR}, written in the form
\begin{eqnarray}
\label{DDR}
|A_2| \leq \sqrt{R}~,
\end{eqnarray}
where  $A_2$ is the usual transverse asymmetry and
$R=\sigma_L/\sigma_T$ is the standard ratio in DIS of the cross section of
longitudinally to transversely polarized off-shell photons.
It reflects a non-trivial
positivity condition one has on the photon-nucleon helicity amplitudes.
By substituting photons for gluons, we found earlier\cite{ST}, that
the similar bound holds for the various matrix elements for
longitudinal gluons in a nucleon \cite{GI}
\begin{eqnarray}
\label{ineq}
|\Delta G_T(x)| \leq \sqrt{1/2G(x)G_L(x)}~.
\end{eqnarray}
However, this bound can be rederived in line with the positivity bound in the
quark case, known as Soffer inequality \cite{S},
 \begin{eqnarray}
|h_1(x)| \leq q_+(x) = {1 \over 2} [q(x) + \Delta q(x)]~,
\label{S}\end{eqnarray}
by making the substitution in Eq.(2), $G(x) \to G_+(x)
= {1 \over 2} [G(x) + \Delta G(x)]$, and providing a stronger
restriction, especially when the gluon helicity distribution $\Delta G(x)$
is small or even negative.
Coming back to the photon case, if $A_1$ denotes the asymmetry with
longitudinally
polarized nucleon, we are led to
\begin{eqnarray}
\label{DDRn}
|A_2| \leq \sqrt{R(1+A_1)/2}~,
\end{eqnarray}
a stronger bound than Eq.(1).
In the present paper we will show that this is really the case,
using a transparent physical approach,
and generalize it to consider each quark flavor separately.
We show that this leads to sensitive tests for parton distribution
and sometimes give a hint about higher twist terms. We will also comment on,
why, we think the weaker bound was used up to now.

We start with the following expressions for
the various photon-nucleon cross-sections in terms of the
matrix elements describing the transition from the state $|H,h>$ of a
nucleon with helicity $h$ and a photon with helicity $H$,
to the unobserved state
 $|X>$
\begin{eqnarray}
\label{def}
\sigma^{\pm}_T=\sum_{X} |<+1/2, \pm 1|X>|^2~,\nonumber \\
\sigma_L=\sum_{X} |<+1/2, 0|X>|^2=\sum_{X}
|<-1/2,0|X>|^2~,
\nonumber \\
\sigma_{LT}=
2 Re \sum_{X} <+1/2,
+1|X> <-1/2, 0|X>~.
\end{eqnarray}
Note that while longitudinal $\sigma_L$ and transverse 
$\sigma_T=\sigma_T^+ +\sigma_T^-$
cross-sections
are symmetric with
respect to the reverse of the nucleon and photon helicities, this
is not the case for the interference term $\sigma_{LT}$. 
The reason is very simple:
the opposite helicities of photon and nucleon correspond to their spins
parallel,
so that the angular momentum of the state  $|X>$ has its maximum value
$3/2$. The amplitude, which could  possibly interfere with it
to produce the transverse asymmetry, should have the same total
angular momentum of the state  $|X>$.
This is however impossible, as
the flip of the one of the helicities
would require another one to exceed its maximal possible value,
in order to keep the angular momentum of $|X>$ the same.
Therefore the interference, responsible for $A_2$, does not occur.
This is quite a general reason,
for the occurence of the $+$ helicity configurations
in all the cases considered above.

We are now ready to write down the Cauchy-Schwarz inequality as
\begin{eqnarray}
\label{CS}
\sum_{X}|<+1/2,+1|X> \pm a<-1/2,0|X>|^2 \geq 0~,
\end{eqnarray}
where $a$ is a positive real number.
By making use of the definitions (\ref{def}) and
after the standard minimization with respect to the choice of
$a$, one immediately arrives at
\begin{eqnarray}
\label{DDRs}
|\sigma_{LT}| \leq \sqrt{\sigma_L \sigma^+_T}~,
\end{eqnarray}
leading directly to (\ref{DDRn}).

The next important step is to apply this method for
each quark flavor separately.
This may be achieved by considering a fictitious ``photon''
coupled to only one flavor. In other words, this is just the consideration
of positivity \cite{DDR}
for each flavor contribution to the structure function $W^{\mu \nu}$, i.e.

\begin{eqnarray}
\label{wf}
W^{\mu \nu}_f \epsilon_\mu \epsilon^{*}_\nu > 0.
\end{eqnarray}
So, if we have the following definition
\begin{eqnarray}
\label{defq}
\sigma_{i} = \sum_f e^2_f \sigma_{i}^f (i=L,T,LT),
\end{eqnarray}
positivity implies, that
\begin{eqnarray}
\label{DDRq}
|\sigma_{LT}^f| \leq \sqrt{\sigma_L^f \sigma^{+f}_T}~,
\end{eqnarray}
or analogously to eq. (\ref{DDRn})
\begin{eqnarray}
\label{DDRnf}
|A_2^f| \leq \sqrt{R^f(1+A_1^f)/2}~.
\end{eqnarray}
Note of course that all the kinematic factors correspond
to the nucleon, so that, say,  no quark masses will appear.

We are now going to make use of the fact that these inequalities
involve terms of different twists which may be used
to put some constraints on their relative size and $Q^2$-evolution
\cite{OT98}. The improved bound at hand is a practical tool
to start such an investigation. Let us consider, as a first step, the
twist-two approximation at Born level
for both $\sigma_L^f$ and $ \sigma^{f}_{LT}$.
This means that we will take the Wandzura-Wilczek (WW)
\cite{WW}approximation for
$ \sigma^{f}_{LT}$ and the target mass approximation \cite{GP} for
$\sigma_L^f$.

Consequently, one should keep only the twist-two part of the
transverse spin structure functin $g_T(x)=g_1(x)+g_2(x)$
for each quark flavor $f$

\begin{eqnarray}
\label{defg2}
g^f_T(x) = g^{f,WW}_T (x) + \bar g^f_2 (x)
\end{eqnarray}
with
\begin{eqnarray}
g^{f,WW}_T (x)=
\int_x^1\frac{dz \Delta f(z)}{z},
\end{eqnarray}
represented by the first term in the r.h.s., disregarding the
twist-three
part, which, in this case, is entirely provided
by the twist-three part of structure
function $g_2$. At the same time, one should disregard the twist-four
part
of the longitudinal
structure function $F_L$, as well as the twist-two part coming
from the radiative corrections \footnote{Note that the gluon distribution will
enter the constraint for each quark flavor at next-to-leading order.}.
The quantities $A_2^f$ and $R^f$ should then be effectively replaced by

\begin{eqnarray}
\label{def2}
A_2^f
\to \frac{2M}{Q}x g_T(x) \to \int_x^1\frac{dz \Delta f(z)}{z}; \nonumber \\
R^f \to \frac{F^f_L(x)}{F_2^f(x)} \to
\frac{4M^2 x^2}{Q^2}\frac{\int_x^1\frac{dz f(z)}{z}}{f(x)}.
\end{eqnarray}
Then positivity requires that
\begin{eqnarray}
\label{in2}
|\int_x^1\frac{dz \Delta f(z)}{z}| \leq
\sqrt{f_+(x) \cdot \int_x^1\frac{dz f(z)}{z}}=
\sqrt{\frac {f(x)+\Delta f(x)}{2}
\cdot \int_x^1\frac{dz f(z)}{z}}.
\end{eqnarray}

We checked this bound and found that it is strongly dependent on the
parametrization one uses for the quark distributions.
Especially sensitive is the case
of $d$ quark, because of its negative polarization.
While the
inequality is satisfied by the GRSV \cite{GRSV} distributions,
the GS \cite{GS} distributions exhibit a tiny
violation of the
inequality in the case of $d$-quark for $0.4 \leq x \leq 0.8$
(see Fig.1), but not for the $u$-quark (Fig. 2). We also anticipate
a violation in the case of the strange quark, whose polarization,
although losely known, is expected to be negative.

\begin{figure}
\begin{center}
\leavevmode {\epsfysize=16cm \epsffile{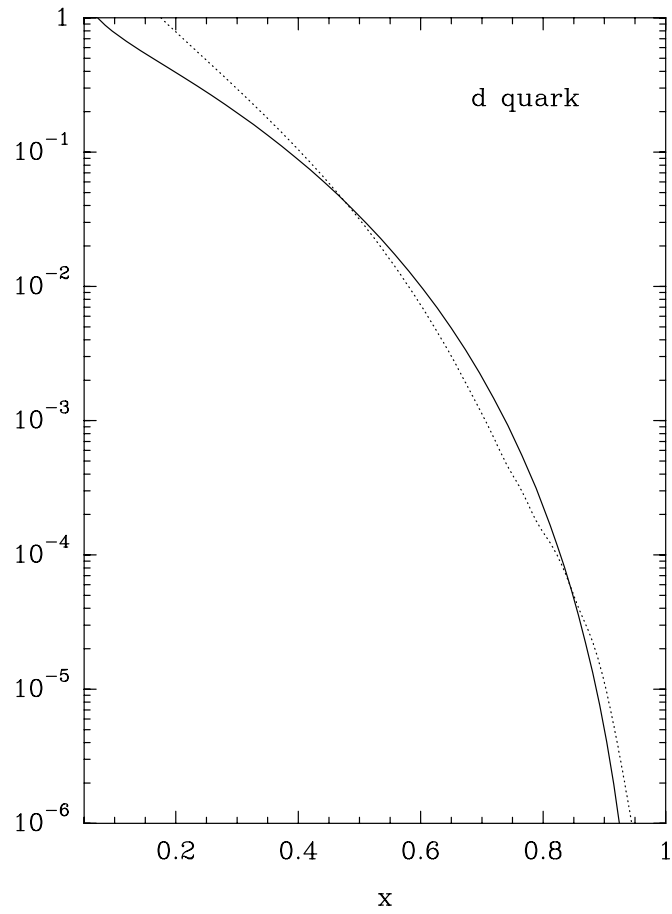}}
\end{center}
\caption[*]{\baselineskip 13pt
Test of the positivity bound using the GS distribution
for $d$ quark at $Q^2=1 GeV^2$.
The solid curve corresponds to the l.h.s. of eq
(\ref{in2}) and the dotted curve to the r.h.s.
}\label{Fig1}
\end{figure}

\begin{figure}
\begin{center}
\leavevmode {\epsfysize=16cm \epsffile{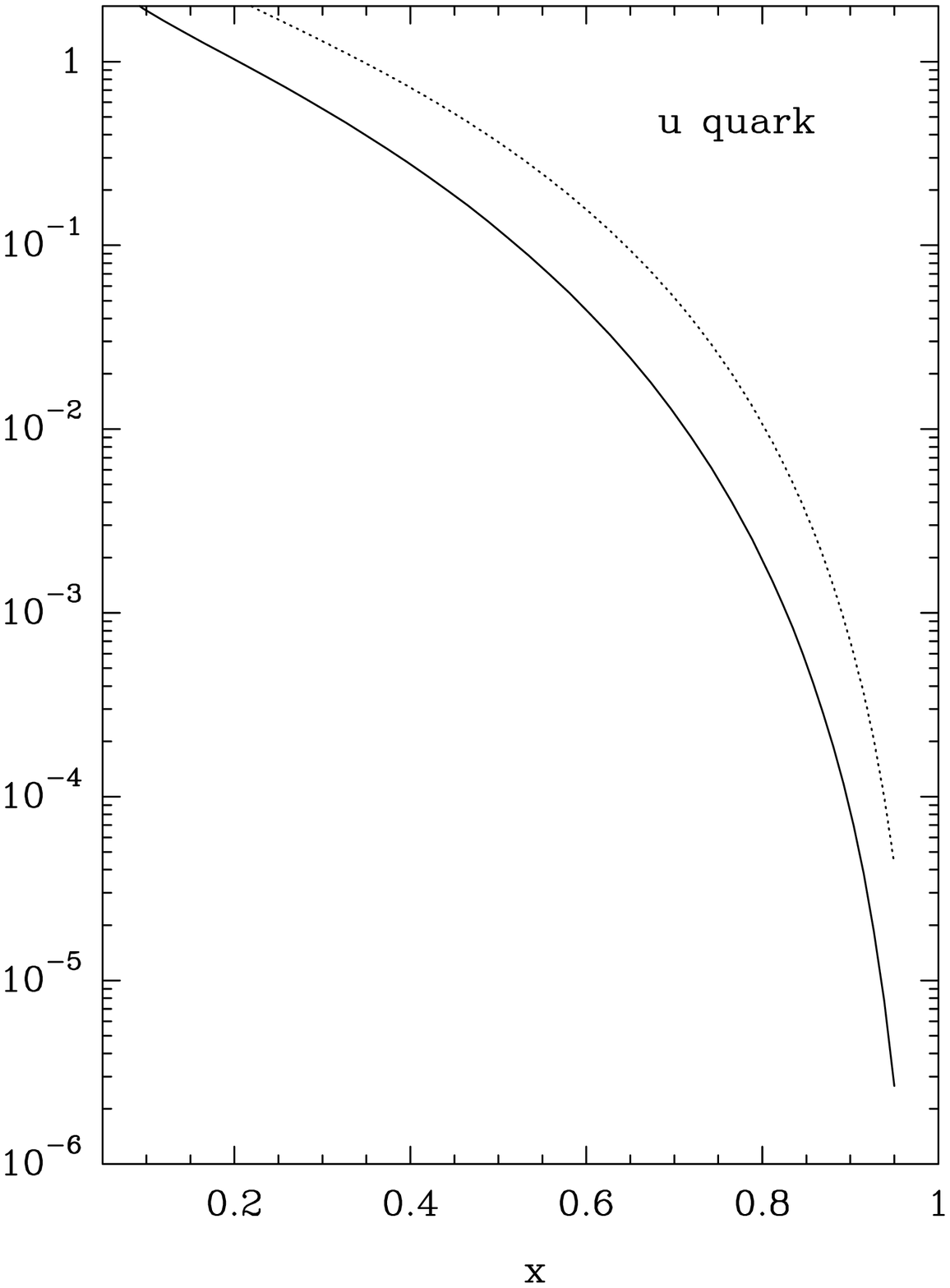}}
\end{center}
\caption[*]{\baselineskip 13pt
Same as Fig.~\ref{Fig1}, but for $u$ quark.

}\label{Fig2}
\end{figure}

We make two general conclusions from such a controversial situation.
First, the generalized bound (\ref{in2}) is a new
sensitive test for the
quark distributions. Second, the precise
experimental data in the large $x$ region, which should be obtained at
CEBAF, will be of great importance to resolve the controversy,

If the violation of the constraint is confirmed, it will be
considered as an indication of the inconsistency of the twist-two
Born approximation and the
necessity to account for higher orders of perturbation theory
and higher twists. The simplest way to do this would be
to assume a non-zero twist-three $\bar g_2^f$ (see eq.
(\ref{defg2})).

This guess is also justified by the fact, that the constraint
(\ref{DDRnf}) may be, in principle, violated by WW approximation
regardless of the approximation one uses for $F_L$.
To see that, let us consider a parton distribution, for which the
quark polarization is large and negative in the region of moderate $x$.
Then, on the one side, $g_T^{WW}$ should be large. At the same time,
the quantity $f_+$ is small and the bound is violated.
Physically, this can be interpreted in the following way. The WW
approximation may be obtained \cite{OT95} by accounting for an
intrinsic transverse momentum $k_T$ of the polarized quark.
Violation of positivity tells, that such a $k_T$ kick,
in general, may result in the hadronic remnant $|X>$,
not being physical.
The approximation for $\sigma_L$ is actually not very important, since
the violation of positivity comes from the smallness of the amplitude
$<+1/2,+1|X>$, which does not enter in the definition of $\sigma_L$.
The key ingredient of WW,
resulting in this property is the expression of $g_T$ entirely in terms
of $g_1$. Therefore, the twist-three contribution is, in general,
a necessary ingredient to restore positivity.

The $d$ quark, which has a negative polarization makes such a
qualitative arguments
more profound. This may lead  to a long-awaited qualitative reason
for the strong flavor dependence of twist-three contributions
found by QCD sum rule method
\cite{BBK,Sh}. It is in fact
resulting in an order of magnitude larger value of its
third moment
\begin{eqnarray}
\label{mom3}
\int_0^1 dx x^2 \bar g_2^f(x) \equiv \frac{1}{3} d^f_2
\end{eqnarray}
for neutron with respect to proton, and consequently for
$d$ quark with respect to $u$ quark.

Although it may be too early to use the current parametrizations
for
quantitative estimates, let us discuss some relevant implications.
Note first, that because the WW contribution is negative for $d$ quark,
the twist-three part should be positive to restore the validity
of the constraint. At the same time, to make any statement
about the sign of $d_2$ one should take into account, that
$\bar g_2$ should change sign due to the Burkhardt-Cottingham sum rule
(whose possible violations cannot appear in the framework of
operator-product expansion, being the basis of quantitative
analysis of twist-three effects) \cite{ST95}. Moreover, if the
sea quark contribution is negligible, the sign changes should be even
more dramatic due to the so-called
Efremov-Leader-Teryaev sum rule\cite{ELT}.
Surprizingly enough, the simple model \cite{OT95},
based on these two sum
rules predicts the existence of a maximum of $\bar g_2$ at $x \sim 0.6$,
which is not far from the region where positivity at twist-two level
is violated. Moreover, the sign of the
third moment happens to be opposite
to the sign of $\bar g_2$ in this point, so that the positive value
required by positivity  corresponds to the negative moment,
like it was obtained by the QCD sum rule method.
At the same time, the value of that moment, due to oscillations, was
predicted to be an order of magnitude smaller than
the value at this point.
The minimal value of $\bar g_2$ for
GS parametrization is about $10^{-3}$.
This corresponds to a bound of the third  moment from below
of about $10^{-4}$, two orders of magnitude smaller than
the QCD sum rule calculations.

One may note several possible reasons for such a discrepancy.

i) More accurate data may increase the degree of the twist-two
positivity violation.

ii) The actual strength of the twist-three contribution may
essentially exceed its
lower positivity bound.

iii) The model \cite{OT95} may be too crude.


iiii) The (relatively) large value of twist-three contribution
for $d$ quark is an ingredient of the
QCD sum rule calculation rather than
of the real nucleon structure. This point of view is supported by the
fact that the recent SLAC data \cite{SLAC}
seem to disagree with the QCD sum rule calculations.

Let us discuss this possibility in more details.
The existence of a large
twist-three contribution is due to the two sources -
positivity constraint
and large negative polarization. The positivity is likely to be present
in sum rule calculations, as their main ingredients are spectral
densities,
corresponding to the on-shell particles in the intermediate states,
which are in turn the main ingredients of the positivity proofs.
At the same time, the
$d$-quark negative polarization is introduced by
local nucleon currents.
As soon as QCD sum rules are unapplicable to describe the
region of large
and small $x$, and their ability to describe the
$x$-dependence of parton
distributions is rather limited, one can guess, that the $d$-quark
polarization may be in fact
overestimated in the region of moderate $x$,
which is crucial to the positivity constraint. If so, this implicit
overestimated polarization would be the actual source of the
strong twist-three contribution,
which should restore the validity of positivity
constraints. One may speak about ``positivity-driven'' twist-three
and it is a matter of future studies to decide, if it does reflect
real physics or if it is just a consequence of the poor
description of the $d$-quark longitudinal polarization.


The use of the new bound is resolving partially the puzzle,
why the measured $A_2$
is such a small quantity. The fact, that the bound (1) is far
from being saturated is obvious at low $x$ in the proton
case, because, according to (4), it should be
decreased by a factor $\sqrt 2$,
due to the small longitudinal asymmetry.
The bound under consideration is even more useful with a negative
longitudinal asymmetry, like in the neutron case.
At the same time, the nucleon bounds are still undersaturated
in comparison with the quark bounds, especially for $d$-quark,
which may be saturated or even violated for some models.
With the improved determination of the unpolarized $s(x)$ and $\bar s(x)$
distributions, positivity can be used to put a non trivial bound on the
$\Delta s(x)$ and $\Delta \bar s(x)$.\footnote{ We are grateful to J.Ellis
for this relevant comment}
The general conclusion is
that the bounds for each flavor are more sensitive, while
their effects may be diluted for nucleons (especially, for proton).

We are indebted to E. Leader,
E. de Rafael, A. Sch\"afer, A. Tkabladze and J.-M. Virey
for useful discussions and correspondence and to
Z-E. Meziani and R. Windmolders for interest in the work.
O.T. is thankful to Centre de Physique Th\'eorique (CNRS-Luminy)
for warm hospitality and to Universit\'e de Provence for
financial support.

\end{document}